\documentclass[aps,prl,twocolumn,groupedaddress,showpacs]{revtex4}
\usepackage{graphicx}

\newcommand{\delete}[1]{}

\newcommand{\be}{\begin{equation}}
\newcommand{\ee}{\end{equation}}

\def\beq{\begin{equation}}
\def\eeq{\end{equation}}
\def\bea{\begin{eqnarray}}
\def\eea{\end{eqnarray}}
\def\ba{\begin{array}}
\def\ea{\end{array}}

\begin{document}

\title{Resonant interaction of trapped cold atoms with a magnetic cantilever tip}
\author{Cris Montoya, Jose Valencia, Andrew A. Geraci}
\email[]{ageraci@unr.edu}
\affiliation{Department of Physics, University of Nevada, Reno, NV 89557}
\author{Matthew Eardley, John Moreland, Leo Hollberg$^1$, John Kitching}
\affiliation{National Institute of Standards and Technology, Boulder, CO 80305}
\thanks{$^1$ present address: Stanford University, Stanford,CA 94305.}

\date{\today}
\begin{abstract}

Magnetic resonance in an ensemble of laser-cooled trapped Rb atoms is excited using a micro-cantilever with a magnetic tip. The cantilever is mounted on a multi-layer chip designed to capture, cool, and magnetically transport cold atoms. The coupling is observed by measuring the loss from a magnetic trap as the oscillating cantilever induces Zeeman state transitions in the atoms. Interfacing cold atoms with mechanical devices could enable probing and manipulating atomic spins with nanometer spatial resolution and single-spin sensitivity, leading to new capabilities in quantum computation, quantum simulation, or precision sensing. 


\end{abstract}


\maketitle




{\it{Introduction.}} Mechanical oscillators, such as micro-cantilevers, have demonstrated remarkable force sensitivity, enabling the detection of single-electron spins in solids \cite{rugar2} and tests for non-Newtonian gravity at sub-millimeter length scales \cite{stanford08}. Cantilevers that utilize strong magnetic gradients can achieve excellent spatial localization, allowing nanoscale magnetic resonance imaging \cite{nanoMRI,nanoMRI2,nanoMRI3}. A mechanical resonator coated with ferromagnetic material was recently used to coherently control the spin state of a single Nitrogen-Vacancy (NV) center in a diamond substrate \cite{lukinNV}, allowing sensitive scanning magnetometry with nanometer spatial resolution. NV-centers have also been used to coherently sense the motion of a magnetized cantilever beam \cite{lukinharris}. Scanning NV magnetometry has been used to detect single electron spins under ambient conditions \cite{lukinNV2}. It has been proposed that mechanical resonators loaded with a magnetic tip could be interfaced with bose-einstein condensates \cite{philippbec,nanowire} or single atoms to implement quantum gates \cite{andyjohnpra}. Coupling of cantilevers to atoms has been demonstrated using magnetic interactions for room-temperature atomic gases \cite{wang} and using surface forces for BECs \cite{treutlein2010}, important steps toward the goal of probing and manipulation of atomic spins with micrometer-scale resolution and single-spin sensitivity.  Here we demonstrate magnetic coupling of a cantilever to an ensemble of cold atoms. This approach provides a different set of advantages and challenges as compared with other techniques involving coupling to spins in solid state systems.

Such mechanical spin exchange coupling can be coherent and non-destructive. Advances in this area could lead to new capabilities in neutral atom quantum computation and quantum simulation, microscopy and magnetic resonance force microscopy, precision force sensing, or atomic clocks that couple mechanics to an atomic resonance \cite{pisano}. For instance, optical lattices can confine regular arrays of atoms where large subsets of atoms can be entangled, making these systems well suited for measurement-based quantum computation schemes \cite{cluster1,cluster2}. 
A scalable method to couple atoms occupying individual lattice sites with an array of cantilevers has been described \cite{andyjohnpra}.  Here each lattice site in a 2-D plane is paired with a magnetic cantilever tip to allow massively parallel single qubit operations. The interaction of the cantilever with the atomic spin can be fully coherent, making this system promising for protocols requiring multiple manipulations on a single qubit.

Mechanical resonators can be used for quantum simulation protocols \cite{qsim1,qsim2,qsim3}, in which arrays of individual neutral atoms are confined in large two-dimensional lattices.  
Using the mechanical approach, both the site population and spin state can in principle be determined, while complications involving light assisted collisions \cite{lightassistcoll} that occur for optical readout methods can be avoided. It is possible to confine atoms using surface plasmonic modes in order to achieve sub-micron lattice spacing, allowing very fast tunneling rates \cite{lukinplasmonic}.  Mechanical resonators could provide high-resolution imaging in such systems, without being limited by optical diffraction.



In this paper, we report on progress toward these goals by demonstrating the excitation of magnetic resonance in an ensemble of laser-cooled trapped Rb atoms using a micro-cantilever with a magnetic tip. The coupling is observed by measuring the loss from a magnetic trap as the oscillating cantilever induces Zeeman state transitions in the atoms.  The cantilever is mounted on a multi-layer atom chip designed to capture, cool, and magnetically transport cold atoms.  We also show that for realistic parameters, mechanical detection of individual atomic spins should be possible in a cryogenic system. 


{\it{Experimental Setup.}} In the experiment, Rb atoms emerge from a dispenser located approximately 5 cm from the chip surface and are collected in a mirror-magneto-optical trap (mirror-MOT) \cite{mirrormot} approximately 2.5 mm above the chip surface. A portion of the chip surface is coated with $500$ nm of Cu to provide a mirror for the MOT beams.  Cu is chosen to minimize adsorption of Rb on the surface \cite{germanpaper}.  Following this initial MOT stage, which employs an external anti-Helmholtz field to generate a trap gradient of $10$ G/cm, the MOT is transferred closer to the chip surface $\sim 1.1$ mm in a U-MOT stage, where the gradient is provided by current in a U-shaped wire embedded in the chip and the atom number is approximately $5 \times 10^{6}$.  Following this stage, the atom cloud is compressed by reducing the hyperfine repump intensity, further cooled to 7 $\mu$K by polarization gradient cooling, and then optically pumped to the $^{87}$Rb $|F=2,m_F=2>$ hyperfine ground state for magnetic trapping.  The atoms are then loaded into a magnetic quadrupole U-trap produced by the same U-shaped wire used to create the U-MOT. Transferring the cold atoms from the U- trap to the cantilevers is facilitated by magnetic guiding above the chip, which involves transport over a distance of approximately $0.8$ cm on a time scale of 200 ms. The guiding field is produced by external magnetic fields and by current running through different wire configurations on the chip. During the transfer process it is necessary to change the axis of the trap to be parallel with the wire guides which lead to the cantilever. This rotation is accomplished with a P-shaped magnetic trap similar to that described in Ref. \cite{ptrap}.

\begin{figure}[!t]
\begin{center}
\includegraphics[width=1.0\columnwidth]{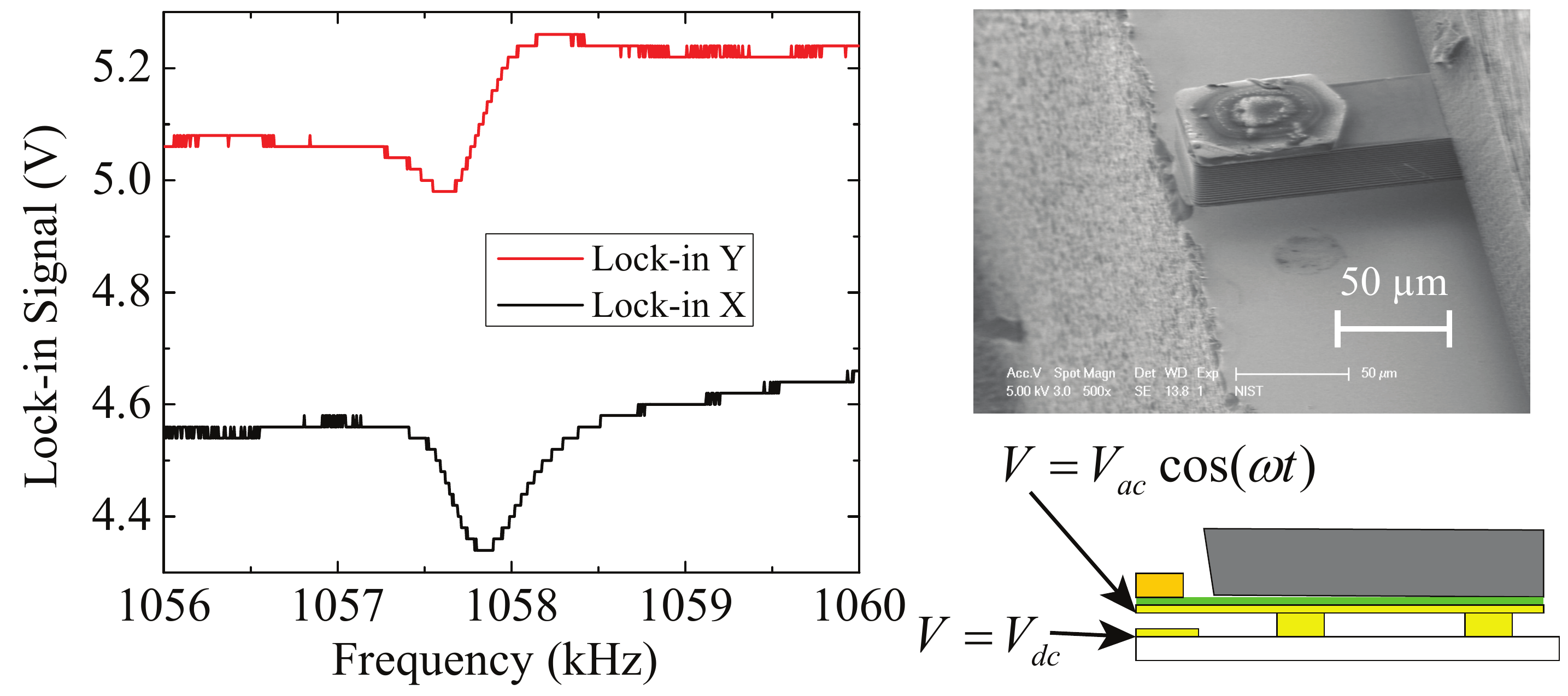}
\caption{(left) Cantilever resonance as measured with a quadrant photodetector and lock-in amplifier. (upper) SEM micrograph of Si cantilever beam with CoNiMnP magnet. (lower) Schematic of capacitive driving mechanism.
\label{setup}}
\end{center}
\end{figure}

The Si cantilevers which are mounted on the atom-chip are shown in Fig. \ref{setup}.  They were fabricated at NIST by patterning and etching a SOI (Silicon-on-Insulator) wafer. They have dimensions of $130$ $\mu$m $\times \:  60$ $\mu$m $\times \: 25$ $\mu$m and are functionalized with an electroplated CoNiMnP \cite{magnets} ferromagnet of dimensions $85$ $\mu$m $\times \: 60$ $\mu$m $\times \: 9$ $\mu$m with a total moment of approximately $2 \times 10^{-9}$ J/T, measured in a commercial magnetometer. The magnet is magnetized with its moment extending outward from the cantilever tip. The measured coercive field is sufficiently high so that we expect the magnet to retain its moment throughout the experiment. The measured $Q$ factor of the cantilevers ranges from $20,000$ to $30,000$ unloaded and is reduced to approximately $10^4$ with the addition of the magnet. This $Q$ factor is consistent with the theoretical limit calculated from thermoelastic dissipation \cite{TED}. The bare cantilevers are annealed in a Nitrogen atmosphere at $700 \;^{\circ}$C for $1$ hour to improve the $Q$-factor. The cantilevers are coated with gold following the annealing procedure, which typially reduces the $Q$ factor to below $10^4$.

 After magnetic transport, the atoms are trapped approximately $100$ $\mu$m from the cantilever tip, in a quadrupole trap formed by the magnetic tip itself and external fields. 
 At this point the atom temperature has increased to  $\sim$100 $\mu$K and the trapped atom number is near $2 \times 10^4$, as determined by absorption imaging.  Once the atoms are trapped near the tip, those atoms in a ``resonant slice'' with a Larmor frequency that matches the cantilever vibrational frequency, will have their spin undergo a coherent precession about the net direction of the total magnetic field, resulting in Zeeman-state transitions.  The dc magnetic field from the cantilever moment combines with the external magnetic field to determine the Larmor precession frequency. The cantilever fundamental-mode resonance frequency when loaded with a magnet of mass $M$ satisfies $\omega_c = \frac{k}{0.24m_c+M}$, for a cantilever of mass $m_c$. Here the spring constant $k=\frac{1}{4} E w (\frac{h}{l})^3$ where $E$ is the Young's modulus and  $l$, $w$, and $h$ are the cantilever length, width, and height, respectively. The measured resonance frequency of the loaded cantilever is $1057.7$ kHz, in reasonable agreement with the expected value.
The Rabi-frequency $\Omega_R$ for driven Zeeman spin-transitions depends linearly on the amplitude of the cantilever motion $\delta z$ and the magnetic gradient $G_m$ of the tip
\be
\Omega_R=\delta z G_m \gamma,
\ee
where $\gamma$ is the gyro-magnetic factor for $^{87}$Rb, $7$Hz / nT. To efficiently observe the influence of the cantilever on the atomic spins, the Rabi frequency should be of order of the average trap frequency $\sim$1 kHz.  For our magnetic moment of $2 \times 10^{-9}$ J/T, the required cantilever amplitude is approximately $50$ nm.

To characterize the motion of the cantilever \emph{in-situ}, a HeNe laser is reflected from the surface of the cantilever onto a quadrant photodetector with a bandwidth of 1 MHz. The quadrant photodetector is positioned on a two-dimensional stage with micrometers to allow a distance calibration. The motion is measured using an RF lock-in amplifier. The cantilever used in these measurements has a mechanical linewidth of $0.67 \pm 0.02$ kHz as determined by a Lorentzian fit to the data shown in Fig. \ref{setup}. The cantilever is driven by applying AC + DC voltage between its metallized under-surface and a nearby electrode. For $V_{dc} = 40$ V, $V_{ac} = 10$ V, and electrode separation $d = 9$ $\mu$m, we measure a tip amplitude of approximately $\delta z = 34 \pm 13$ nm at $1057.7$ kHz, sufficient to generate Rabi frequencies exceeding $100$ Hz. The uncertainty in $\delta z$ is dominated by the finite laser waist size at the cantilever tip, which is comparable to the cantilever width. Reduced uncertainty could be achieved by using a tighter focus or a fiber-coupled laser interferometer. Approximating the system as a parallel plate capacitor, the expected amplitude is given by $\delta z = \frac{Q}{k} \frac{\epsilon_0 A V_{\rm{dc}} V_{\rm{ac}}}{d^2}$, where $A$ is the effective area of the plates. Taking the measured $Q$ factor we calculate an expected tip displacement of $\delta z = 40$ nm, which agrees with the measured result.

The effect of the cantilever motion on the atomic spins in the quadrupole trap can be more rigorously calculated using Landau-Zener theory \cite{LZ1,LZ2}. For example, atoms beginning in the state $|F = 2, m_F = 2>$ will experience a transition into the level $|F = 2, m_F = 1>$ with a probability $P=1-\exp{[-\frac{\pi}{4}\frac{\mu_B B'}{\hbar}\frac{(\delta z)^2}{v}]}$
as they traverse the resonant slice with velocity $v$. Here $B'$ is the magnetic field gradient due to the cantilever tip, of order $10$ T/m at the resonant slice nearest to the cantilever. We model the magnetic field of the magnet assuming a rectangular geometry. One dimensional numerical simulations indicate that for a thermal ensemble at $100$ $\mu$K, population loss due to the transition into strong magnetic field seeking states can be expected over a time scale of $\sim$10 ms for our experimental parameters \cite{mattthesis}.

\begin{figure}[!t]
\begin{center}
\includegraphics[width=0.93\columnwidth]{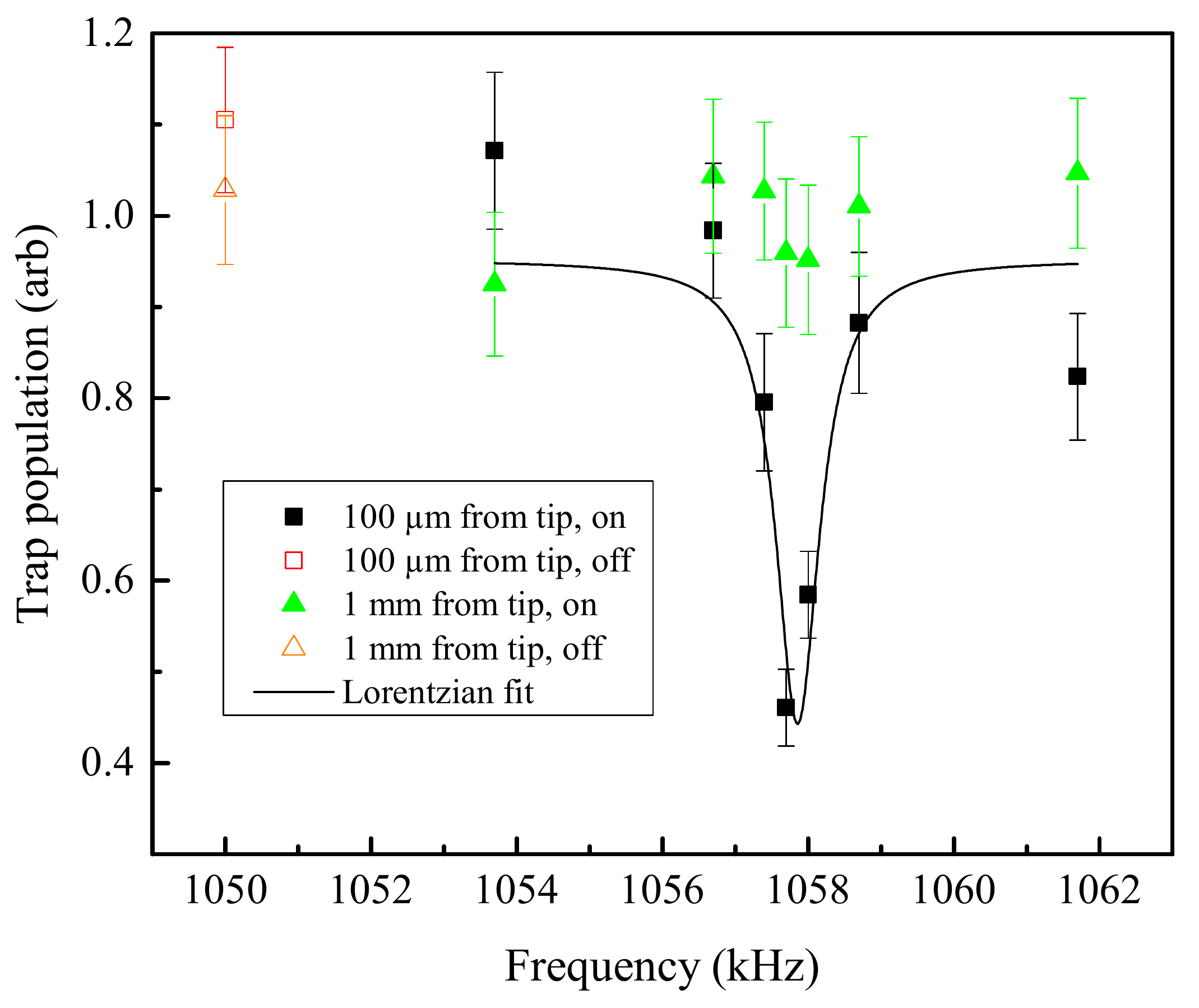}
\caption{Atomic population in the trap determined by absorption imaging, after interacting with the cantilever for $11$ ms at a distance of approximately $100$ $\mu$m (black points). When the cantilever is driven near its resonance of 1057.7 kHz, the trap population decreases (black points). The open points indicate the population with the cantilever not capacitively driven. A Lorenztian fit to the peak is shown, which agrees with the measured linewidth of the cantilever resonance to within error bars. Also shown are data for the atoms separated from the cantilever by a larger distance of $1$ mm (green triangles). As expected, no significant reduction in trap population occurs in this case.
\label{atomcantfig}}
\end{center}
\end{figure}

{\it{Results}}.  The interaction with the cantilever becomes significant when the atomic cloud is moved to approximately $100$ $\mu$m from the oscillating cantilever tip. After a variable time $t$ of interaction, a low intensity laser is used to probe the trap population via absorption imaging. A frame-transfer Peltier-cooled CCD camera is used for imaging, in a differential imaging mode. A hyperfine depumping pulse is applied between the images which optically pumps atoms into the $F=1$ state which is a dark state for the imaging light. Fig \ref{atomcantfig} shows the trap population after $t=11$ ms of interaction with the cantilever driven with $V_{\rm{ac}}=10$ V and $V_{\rm{dc}}=40$ V. We define time $t=0$ as the approximate time that the atomic cloud centroid approaches $100$ $\mu$m from the tip. The population remains relatively constant when the cantilever is not driven or driven off-resonance. When the cantilever is driven on resonance 
the population gets reduced by approximately half. 
The population reduction is consistent with one-dimensional Landau-Zener simulations, to within the uncertainty of the measurements. A Lorenztian fit of the peak with fitted width $0.73 \pm 0.3$ kHz agrees with the measured linewidth of the cantilever resonance. With the cloud trapped at larger separation from the cantilever (approximately 1 mm) we observe no significant reduction in the trap population, as expected, due to the steep reduction in Rabi frequency at larger distances. 

\begin{figure}[!t]
\begin{center}
\includegraphics[width=0.9\columnwidth]{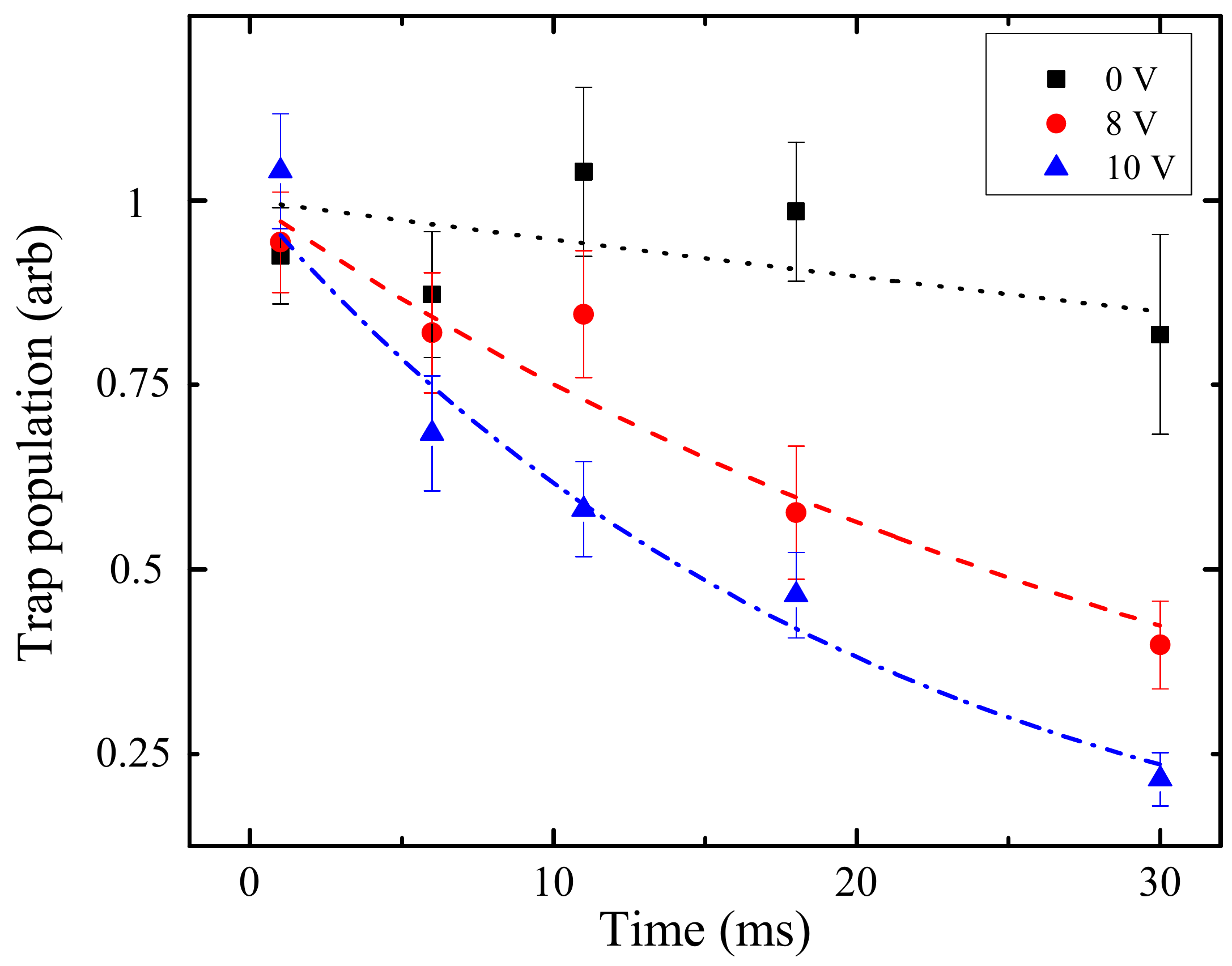}
\caption{Relative trap population after atomic cloud has interacted with the cantilever for varying times, for two different amplitudes of cantilever motion ($8$ V and $10$ V) and with the cantilever not driven ($0$ V). Also shown are fitted exponential decay curves for each case.
\label{signalvtime}}
\end{center}
\end{figure}

In Fig. \ref{signalvtime} we show the trap population after interacting with the cantilever for a varying time, at a distance of approximately $100$ $\mu$m. Results are shown for the cantilever driven with $V_{\rm{ac}} = 0,8$, and $10$ V (with $V_{dc}$ constant at 40V). The results again are suggestive of trap loss due to magnetic Zeeman transitions. We expect the transition probability to vary quadratically with the cantilever driving voltage over this range of interaction strength.  Considering the calculated probability of undergoing a spin flip per pass of the atom through the resonant slice, and the approximate expected rate of crossing due to the atomic motion, the data is in reasonable qualitative agreement with theoretical expectations from the 1-dimensional Landau-Zener model. The fitted observed trap population decay times of $21 \pm 3.1$ ms and $34 \pm 3.6$ ms agree with the $1$-D model predictions to approximately within a factor of two. As the actual atomic motion in the trap occurs in 3-dimensions and can be quite complex, we do not expect exact quantitative agreement for the 1-dimensional calculation. We also include an exponential decay fit to the case with $V_{\rm{ac}}=0$ which yields a significantly longer time constant $184 \pm 66$ ms, in agreement with the measured magnetic trap lifetime.

\begin{figure}[!t]
\begin{center}
\includegraphics[width=1.0\columnwidth]{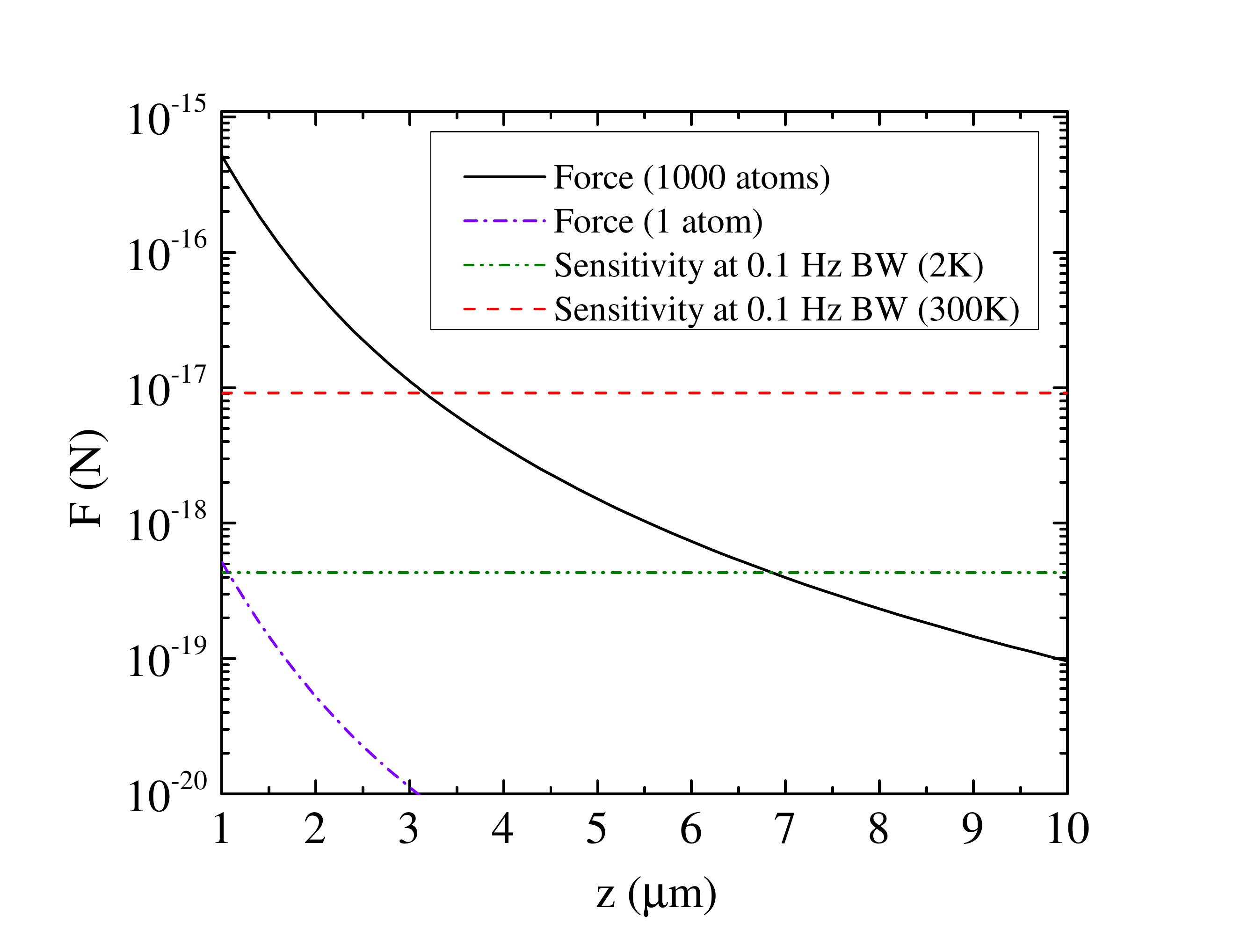}
\caption{Thermal noise limited force sensitivity in a $0.1$ Hz bandwidth for a cantilever as described in the text, and force due to the precession of $10^3$ atomic spins (solid line) and a single spin (dashed line) as a function of separation distance $z$ from the center of the cantilever magnet.
\label{forcefig}}
\end{center}
\end{figure}

{\it{Spin Detection.}} In this work we have demonstrated the ability to induce Zeeman state transitions in trapped cold atoms using the oscillating cantilever tip. It is also interesting to consider detecting the precessing spins by observing the force they impart on the cantilever. A requirement for detecting the back-action of the atomic spins onto the cantilever is that the force induced exceeds the thermal noise-limited force sensitivity of the cantilever. The minimum detectable force due to the presence of thermal noise in a cantilever beam with natural frequency $\omega_c$ can be expressed as
\be
F_{\rm{min}}  = [4 k k_B T b/ \omega_c Q ]^{1/2},
\label{fmin}
\ee
where $b$ is the bandwidth of the measurement and $T$ is the effective temperature of the mode under consideration. The cantilevers used in this work carry a large magnet to facilitate easier atomic trapping, and have not been optimized for detecting the force from atomic spins on the cantilever.  To investigate spin-detection, we can utilize smaller, lower $k$ cantilevers to allow larger cantilever displacements. For example, with a SiN cantilever with dimensions $50$ $\mu$m $\times \: 1.1$ $\mu$m $\times \: 0.2$ $\mu$m and loaded with a CoNiMnP electroplated magnet of dimensions $1.1$ $\mu$m $\times \: 0.9$ $\mu$m $\times \: 0.7$ $\mu$m, the resonance frequency is $\omega_c=2\pi \times 70$ kHz. 
Assuming $Q =10^5$, the sensitivity at room temperature of such a cantilever is $\sim$29 aN$/{\rm{Hz}}^{1/2}$, enabling the detection of $80$ atomic (electron) spins at $1.3$ $\mu$m separation in a $1$ Hz bandwidth. Such small separation from a surface could be realized with an optical trap \cite{lukinphotonic}. For cryogenic cantilevers at $2$ K, the $Q$ factor can be expected to increase.  Si cantilevers of similar size have been demonstrated with Q factors exceeding 380,000 \cite{degen}.  Assuming $Q =$ 300,000 at $2$ K the force sensitivity becomes $1.3$ aN$/{\rm{Hz}}^{1/2}$, allowing single-electron spin detection in a $\sim$0.1 Hz bandwidth at $1.1$ $\mu$m tip-atom separation, as shown in Fig. \ref{forcefig}. 
While demonstrated spin coherence times have been long in surface traps \cite{coherencechip}, at these distances the Casimir-Polder potential needs to be considered, as well as magnetic field noise due to thermal currents in nearby conducting surfaces \cite{CasimirPolder,johnsonnoise}. For atom-surface separations of greater than $0.5$ $\mu$m these effects can be mitigated for suitable material thicknesses and conductivities \cite{andyjohnpra}.

{\it{Discussion.}} We have demonstrated the manipulation of the hyperfine Zeeman states of magnetically trapped $^{87}$Rb atoms using a driven cantilever beam functionalized with a magnetic tip. With optically trapped atoms and improved imaging, coherent control of single trapped atomic spins should be possible in a similar setup. Using a suitably scaled cryogenic setup, the spin of cold atoms can be sensed by measuring the force they exert on a micro-cantilever, with the possibility of single-spin sensitivity for realistic experimental parameters. Such cantilevers could be useful for quantum simulators in which arrays of individual neutral atoms are confined in large two-dimensional lattices. 
It has been shown theoretically that magnetic cantilevers can couple to ultra-cold atoms at $\mu$m distances in a regime analogous to the strong coupling regime of cavity quantum electrodynamics (QED) \cite{philippbec}.  If the mechanical oscillators can be cooled to their quantum ground state \cite{cleland,teufel2,painter}, non-classical states of the atomic degrees of freedom can be transferred to the motional states of the resonators and vice versa, with applications in quantum information science \cite{andyjohnpra,philippnew,law}. In these setups, mechanical oscillators can provide coupling between photons, spins, and charges via phonons \cite{transducers1,transducers2,transducers3,transducers4}. Such transducers could be useful for quantum networks, by allowing coupling between different types of quantum systems, each with different advantages.

{\it{Acknowledgements.}} The authors thank Y.-J. Wang for experimental assistance at the early stages of this work. We thank Leo Hollberg, Philipp Treutlein, and Jonathan Weinstein for useful discussions.  This work was supported in its early stages by DARPA. AG is supported in part by grant NSF-PHY 1205994. This work is a partial contribution of NIST, and is therefore not subject to copyright in the United States.


\begin{thebibliography}{99}

\bibitem{rugar2} D. Rugar, R. Budakian, H.J. Mamin, B.W. Chui, Nature {\bf{430}}, 329 (2004).

\bibitem{stanford08} A. A. Geraci, S. J. Smullin, D. M. Weld, J. Chiaverini, and A. Kapitulnik, Phys. Rev. D {\bf{78}}, 022002 (2008).

\bibitem{nanoMRI} C.L. Degen, M. Poggio, H.J. Mamin, C.T. Rettner, D. Rugar
PNAS {\bf{106}}, 1313  (2009).

\bibitem{nanoMRI2} H. J. Mamin,M. Kim,M. H. Sherwood,C. T. Rettner,K. Ohno,D. D. Awschalom,D. Rugar, Science {\bf{339}} 557 (2013)

\bibitem{nanoMRI3} M. Loretz,T. Rosskopf,J. M. Boss, S. Pezzagna,J. Meijer,C. L. Degen, Science 10.1126/science.1259464 (2014).

\bibitem{lukinNV} S. Hong, M.S. Grinolds, P. Maletinksy, R.L.Walsworth, M.D. Lukin, and A. Yacoby, Nano Lett. {\bf{12}},3920(2012).

\bibitem{lukinharris} S. Kolkowitz, A. C. Bleszynski Jayich, Q. P. Unterreithmeier, S. D. Bennett, P. Rabl, J. G. E. Harris, M. D. Lukin, Science {\bf{335}}, 1603 (2012).

\bibitem{lukinNV2} M. S. Grinolds, S. Hong, P. Maletinsky, L. Luan, M. D. Lukin, R. L. Walsworth, and A. Yacoby, Nature Physics  {\bf{9}}, 215 (2013).



\bibitem{philippbec} P. Treutlein, D. Hunger, S. Camerer, T. W. Hansch, and J. Reichel, Phys. Rev. Lett. {\bf{99}}, 140403 (2007).

\bibitem{nanowire} Z. Darazs, Z. Kurucz, O. Kalman, T. Kiss, J. Fortagh, and P. Domokos, Phys. Rev. Lett. {\bf{112}}, 133603 (2014).

\bibitem{andyjohnpra} A.A. Geraci and J. Kitching, Phys. Rev. A {\bf{80}}, 032317 (2009).

\bibitem{wang} Y.-J. Wang, M. Eardley, S. Knappe, J. Moreland, L. Hollberg, and J. Kitching, Phys. Rev. Lett. {\bf{97}}, 227602 (2006).

\bibitem{treutlein2010} D. Hunger, S. Camerer, T. W. Hansch, D. Konig, J. P. Kotthaus, J. Reichel, P. Treutlein, Phys. Rev. Lett. {\bf{104}}, 143002 (2010).


\bibitem{pisano} C.D. White, G. Piazza, P.J. Stephanou, A.P.Pisano, Sensors and Actuators A {\bf{134}} 239 (2007).

\bibitem{cluster1} G. K. Brennen, C. M. Caves, P. S. Jessen, and I. H. Deutsch, Phys. Rev. Lett. {\bf{82}} 1060 (1999).
\bibitem{cluster2} O. Mandel, M. Greiner, A. Widera, T. Rom, T. W. Hansch, I. Bloch, Nature {\bf{425}}, 937 (2003).


\bibitem{qsim1}M.Lewenstein, A. Sanpera, V. Ahufinger, B. Damski, A. Sen(De) and U. Sen, Adv. Phys. {\bf{56}}, 243-379 (2007).

\bibitem{qsim2}I.Bloch, J.Dalibard, W.Zwerger, Rev. Mod. Phys.
{\bf{80}}, 885-964 (2008).

\bibitem{qsim3}W. S. Bakr, J. I. Gillen, A. Peng, S. Folling, M. Greiner, Nature {\bf{462}}, 74-77 (2009).

\bibitem{lightassistcoll} W. S. Bakr, P. M. Preiss, M. Eric Tai, R. Ma, J. Simon, M. Greiner, Nature {\bf{480}}, 500 (2011).

\bibitem{lukinplasmonic} M. Gullans, T. G. Tiecke, D. E. Chang, J. Feist, J. D. Thompson, J. I. Cirac, P. Zoller, and M. D. Lukin, Phys. Rev. Lett. {\bf{109}}, 235309 (2012).

\bibitem{mirrormot} J. Reichel, W. Hansel, and T. W. Hansch, Phys. Rev. Lett. {\bf{83}}, 3398 (1999).

\bibitem{germanpaper} G. Biedermann, private communication.

\bibitem{ptrap} R. Long,T. Rom,W. Hansel,T. W. Hansch,J. Reichel, The European Physical Journal D {\bf{35}}, 125 (2005).

\bibitem{magnets} H. J. Cho and C. H. Ahn, Journal of Microelectromechanical Systems {\bf{11}}, 78 (2002).

\bibitem{TED} K. Y. Yasumura, T. D. Stowe, E. M. Chow, T. Pfafman, T. W. Kenny, B. C. Stipe,
and D. Rugar, JMEMS. {\bf{9}}, 117, (2000).

\bibitem{LZ1} C. Zener, Proc. Royal Soc. London, Series A, {\bf{137}}, 696 (1932).

\bibitem{LZ2} J. R. Rubbmark, M. M. Kashm, M. G. Littman, D. Kleppner, Phys. Rev. A {\bf{23}}, 3107 (1981),  W. Ketterle and N.J. van Druten, Adv. Atomic Molecular and Optical Physics {\bf{37}}, 181 (1996).

\bibitem{mattthesis} M. D. Eardley, Ph.D. thesis, in preparation


\bibitem{lukinphotonic} J. D. Thompson, T. G. Tiecke, N. P. de Leon, J. Feist, A. V. Akimov, M. Gullans, A.S. Zibrov, V.
Vuletic, M. D. Lukin, 10.1126/science.1237125, Science (2013).

\bibitem{degen} C. Degen, private communication.

\bibitem{coherencechip} P. Treutlein, P. Hommelhoff, T. Steinmetz, T. W. Hansch, and J. Reichel, Phys. Rev. Lett. {\bf{92}}, 203005 (2004).

\bibitem{CasimirPolder} Y. Lin, I. Teper, C.Chin, and V. Vuletic, Phys. Rev. Lett. {\bf{92}}, 050404 (2004).

\bibitem{johnsonnoise} D.M. Harber, J.M. Obrecht, J.M. McGuirk, and E.A. Cornell, Physical Review A {\bf{72}},
033610 (2005).


















































\bibitem{cleland} A.D. O'Connell, M. Hofheinz, M. Ansmann, R.C. Bialczak, M. Lenander, E. Lucero, M. Neeley, D. Sank, H. Wang, M. Weides, J. Wenner, J.M. Martinis, A.N. Cleland, Nature {\bf{464}}, 697-703 (2010).

 \bibitem{teufel2} J.D. Teufel, T. Donner, Dale Li, J.W. Harlow, M.S. Allman, K. Cicak, A.J. Sirois, J.D. Whittaker, K.W. Lehnert and R.W. Simmonds, Nature {\bf{475}},359 (2011).

\bibitem{painter} J. Chan, T. P. Mayer Alegre, A. H. Safavi-Naeini, J. T. Hill, A. Krause, S.Gr�blacher, M. Aspelmeyer, and O.Painter, Nature {\bf{478}}, 89, (2011).

\bibitem{philippnew} A. Jockel, A. Faber, T. Kampschulte, M. Korppi, M. T. Rakher, and P. Treutlein, Nature Nanotechnology, doi:10.1038/nnano.2014.278 (2014).

\bibitem{law} C.K.Law and J.H.Eberly, Phys. Rev. Lett. {\bf{76}}, 1055 (1996).

\bibitem{transducers1} K. Stannigel, P. Rabl, A. S. Sorensen, P. Zoller, and  M.D. Lukin, Phys. Rev. Lett. {\bf{105}}, 220501 (2010).

\bibitem{transducers2} C.A. Regal and K. W. Lehnert, J. Phys. Conf. Ser. {\bf{264}}, 012025 (2011).

\bibitem{transducers3} A.H. Safavi-Naeini and O. Painter, New Journal of Physics, {\bf{13}}, 013017 (2011). P. Rabl, S. J. Kolkowitz, F. H. L. Koppens, J. G. E. Harris, P. Zoller and M. D. Lukin, Nature Physics {\bf{6}}, 602 (2010).

\bibitem{transducers4} E. Verhagen, S. Deleglise, S. Weis, A. Schliesser, T. J. Kippenberg, Nature {\bf{482}}, 63 (2012).































































\end{thebibliography}
\end{document}